# Effect of X-Ray Irradiation on Magnetocaloric Material, $(MnNiSi)_{1-x}(Fe_2Ge)_x$

John Peter J. Nunez*[1], Vaibhav Sharma[1], Jessika V. Rojas[1], Radhika Barua[1], Ravi L. Hadimani*[1,2,3]

[1]Department of Mechanical and Nuclear Engineering, Virginia Commonwealth University, Richmond, VA 23284, USA

[2]Department of Biomedical Engineering, Virginia Commonwealth University, Richmond, VA 23284, USA

[3]Department of Psychiatry, Harvard Medical School, Harvard University, Boston, MA 02115, USA

## Abstract

$(MnNiSi)_{1-x}(Fe_2Ge)_x$ composition (x=0.34) alloy was prepared by arc melting, crushed, and sieved to approximately <32 microns. They were utilized in examining the possible magnetic and structural changes when exposed to a dosage of a continuous sweeping rate of ~>120 Gy/min and an absorbed dose of 35 kGy of X-ray radiation. This study reports observable trends in X-ray diffraction and magnetic measurements. Magnetization in the magnetization vs. temperature (both heating and cooling) measurements showed an increase from 2.72 emu/g to 4.01 emu/g in the irradiated sample. The Magnetization vs. magnetic field loops exhibited irradiation-induced magnetic hysteresis. The irradiated sample also exhibited an observable change in its coercivity of $\Delta H_c$ = 14.7% at 200 K. A maximum entropy change $\Delta S_{mag}$ of ~ 11.139 J/kg*K and a $T_{ave}$ peak of 317.5 K was achieved for the pristine sample in comparison to $\Delta S_{mag}$ of ~ 11.349 J/kg*K and a $T_{ave}$ peak of 312.5 K for the irradiated sample. These presented results provided deeper insights into tuning the effect of irradiation to the magnetic properties of $(MnNiSi)_{1-x}(Fe_2Ge)_x$ for composition (x=0.34) that can be utilized in a wide range of magnetocaloric applications in high-energy radiation environments. The irradiation applied to the sample did not induce any structural or magnetic phase changes in the selected composition but modified the magnetic properties marginally.

Corresponding Authors:
John Peter J. Nunez – nunezjj2@vcu.edu
Ravi L. Hadimani – rhadimani@vcu.edu

**Introduction**

Magnetic refrigeration based on magnetocaloric effect (MCE) at room temperature is considered a promising substitution for classical vapor compression systems due to its high efficiency and environmental friendliness. Various materials systems and their magnetocaloric properties have been studied and reported in the literature [1–7]. Although rare-earth-based magnetocaloric materials such as $Gd_5(Si_xGe_{1-x})_4$ exhibit remarkably high isothermal entropy change near room temperature, they are expensive and need complex and prolonged thermal treatment procedures [8–11]. Some of these materias have been studied for biomedical applicatiosn due to their extreme magnetic property changes at transition temperatures [12-15]. $(MnNiSi)_{1-x}(Fe_2Ge)_x$ belongs to the MTX system where M and T are transition metals such as Mn, Fe, Ni, and Co and X is a p-block element such as Si, Ge, Ga or Sn are attractive multicaloric materials that change their magnetic properties with the application of magnetic field, pressure, and temperature [6,16,17]. In addition to their potential refrigeration and cooling effect, they exhibit large magnetostriction/volumetric strain upon application of a magnetic field and pressure near room temperature. These properties make this material system attractive for sensor and actuator applications. The transition temperature of the MTX system on which the material system this study utilized most, specifically the $X = 0.34$ composition can be tuned by varying the composition of constituent materials. MnTX (T = Co, Fe, Ni; X = Si, Ge), the composition utilized in this study, $x = 0.34$ exhibits a first-order magnetic phase transition from a paramagnetic (PM) to a ferromagnetically ordered state and a structural transition from a high-temperature $Ni_2In$-type hexagonal phase (P63/mmc) to a low-temperature TiNiSi-type orthorhombic structure (Pnma) with the hexagonal phase having the smallest unit cell volume [6]. By controlling the composition by doping, or isostructural substitution, it is possible to bring the two transitions close together so that the magnetic and structural transitions become coupled hence, designing magnetocaloric materials to a specific temperature range. Furthermore, at this magneto-structural transition, there is a large decrease in the volume of the unit cell (3%–4%), and the transition temperature becomes highly sensitive to the application of hydrostatic pressure.

Materials that undergo a first-order phase transition and or high accompanied strain will experience training effects due to the generation of micro-cracks or a substantial number of dislocations. These defects progressively increase and deteriorate magnetocaloric and strain effects over the life span of the material operation. In addition to defects formed due to large strain from the first-order phase transition, materials also experience the generation of defects from ionizing radiation. Several researchers have reported significant effects of various types of radiation magnetocaloric materials [18–20]. The effect of proton irradiation in magnetocaloric material, $La(Fe_{0.89}Si_{0.11})_{13}$ resulted in an increase in atomic displacement as the increase of lattice constant and Curie temperature. Moreover, the maximum hysteresis loss almost disappeared, and the relative cooling power increased slightly under proton irradiation [19]. Nanocrystalline materials should present a better irradiation resistance compared with large-grain materials because the large volume fraction of grain boundaries acts as an important aspect for irradiation-induced defects [21]. Assar et al. have focused more on gamma-ray irradiation techniques where the magnetic materials studied were Mg-Cu-Zn and Ni-Cu-Zn ferrites. The authors have shown that gamma-ray irradiation on the Mg-Cu-Zn increased the saturation and remnant magnetization [22]. Mane et al. reported that laser irradiation on polycrystalline cobalt ferrite resulted in an increase in saturation magnetization as laser dose increased in hysteresis loops [23]. In this paper, we investigated primarily on x-ray radiation effect on $(MnNiSi)_{1-x}(Fe_2Ge)_x$ for composition x= 0.34 as this composition has a $T_c$ at approximately room temperature. We have shown that x-ray irradiation did change the magnetization, saturation magnetization permeability, and coercivity but did not affect the isothermal entropy change at a field change of 3T.

**Experimental Procedure**

*Material Synthesis*: Polycrystalline $(MnNiSi)_{1-x}(Fe_2Ge)_x$ (x=0.34) samples were prepared by arc-melting the constituent elements of purity better than 99.9% in an ultrahigh purity argon atmosphere. The mass of all the elements was selected based on the nominal stoichiometric composition. Arc-melting was conducted several times to ensure sample homogeneity. The stoichiometric composition of X = 0.34 was used for this study and additional details on the preparation of this magnetocaloric material are found in Clifford et al. [6] and Deepak et al. [7].

*X-ray Irradiation*: The experimental procedure for this study utilized X-ray radiation from X-RAD 225XL Precision X-Ray with a tungsten target set at 225 kV, 13.3 mA. As shown in Fig. 1, The sample was irradiated with a continuous sweeping rate of ~ >120 Gy/min. X-rays were emitted from the x-ray source in a cone beam with a beam angle of 40 degrees. The uniformity of the beam changes significantly from the outer perimeter thus it collimated the beam angle to 31.4 degrees to achieve a homogenous x-ray beam across the field of radiation. The study proposes that there is an observable effect in the $(MnNiSi)_{1-x}Fe_2Ge_x$ system particularly, for the composition x = 0.34 composition when exposed to an intense amount of X-ray radiation absorbed dose up to ~35kGy.

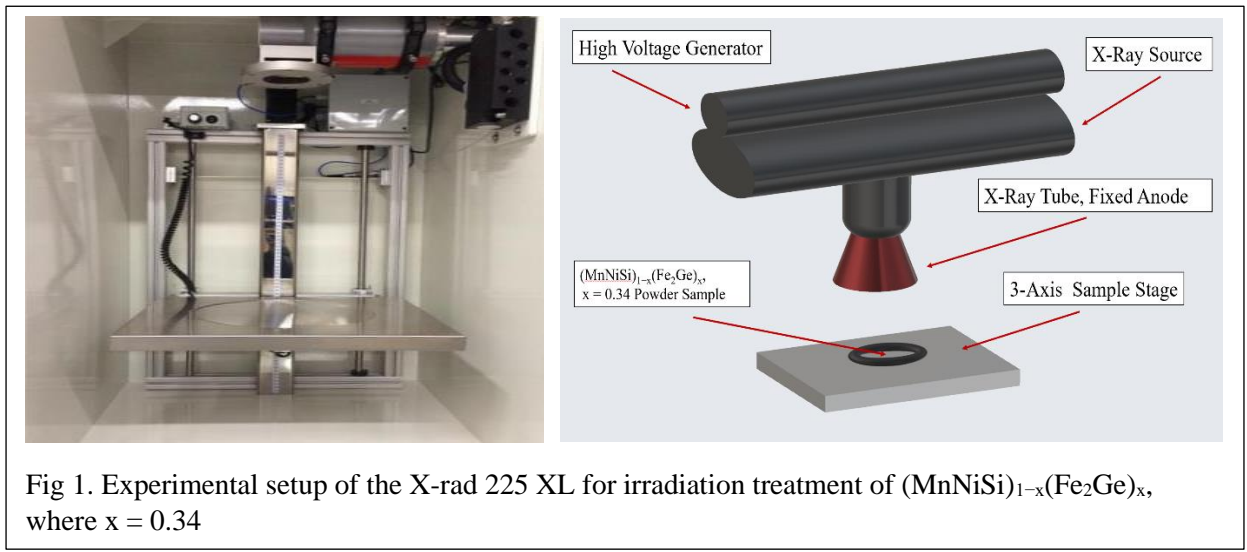

Fig 1. Experimental setup of the X-rad 225 XL for irradiation treatment of $(MnNiSi)_{1-x}(Fe_2Ge)_x$, where x = 0.34

***XRD***: Multipurpose X-Ray Diffractometer using a Copper Kα X-Ray radiation source ($\lambda = 1.54060$ Å) in determining the crystalline structure of the $(MnNiSi)_{1-x}(Fe_2Ge)_x$, where x = 0.34 and any changes associated to irradiation. For this purpose, powders of non-irradiated and irradiated materials were placed onto a silicon wafer in the spinner stage of the X-ray Diffractometer.

***Magnetic Characterization***: A Quantum Design physical property measurement system (PPMS) was used to measure the magnetization (M) of the $(MnNiSi)_{1-x}(Fe_2Ge)_x$ samples within the temperature interval of 200–345 K and, depending on the analysis, within the applied magnetic fields from -3T to 3T.

***Isothermal Entropy Calculations***: The Curie temperature, $T_c$ was determined as the inflection point of the magnetothermal data collected upon heating at a sweep rate of 5 K/min where the derivative with the lowest value between magnetization and temperature was calculated. The magnetocaloric response, entropy change $\Delta S_{mag}$ of the pristine and irradiated samples was calculated by Maxwell relationship applied to the specified isotherms measured at an interval of 2.5 K near the transition temperature and 5 K intervals away from the transition temperature under an applied field of $0T \leq H_{app} \leq 3T$.

**Results and Discussions:** XRD and magnetic measurement results including M vs. T, M vs. H, hysteresis, and ΔS vs. T are presented in this section.

**XRD Analysis:**

The XRD patterns for the pre-and post-irradiated samples are shown in Fig. 2. Their crystalline structure was found to be a hexagonal $Ni_2In$-type structure ($P6_3/mmc$ space group) and orthorhombic (Pnma) TiNiSi-type crystal structure. The patterns of the irradiated samples show slight changes to their peak intensity ratio but do not show peak shifting compared to non-irradiated, which indicates little changes in the interplanar distances. Furthermore, the peaks did not reveal significant changes in the full-width half maximum (FWHM) after sample irradiation.

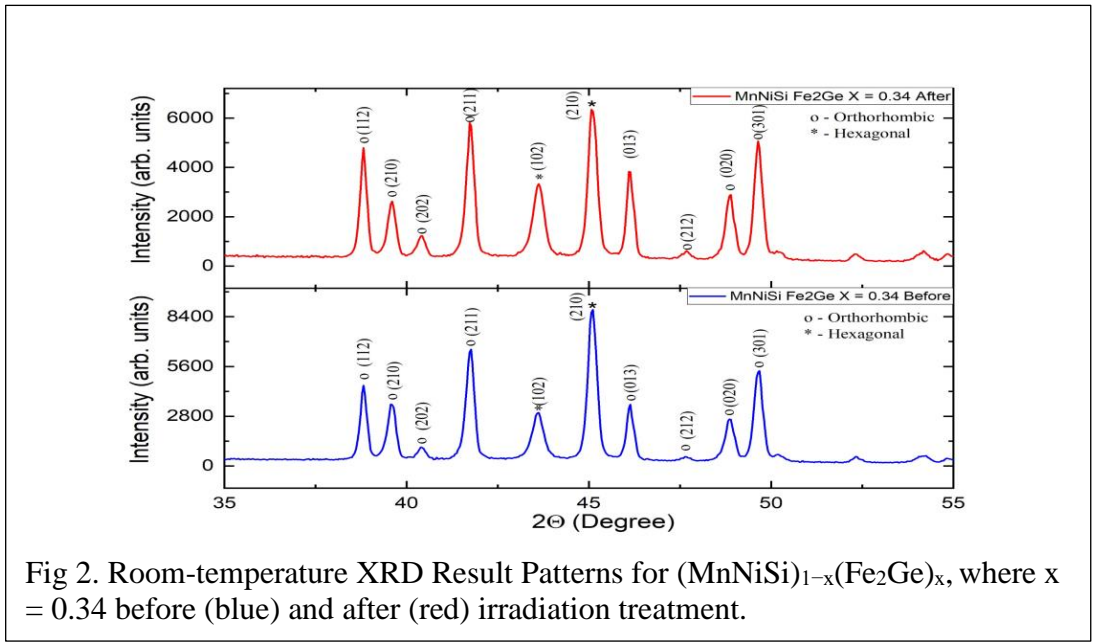

Fig 2. Room-temperature XRD Result Patterns for $(MnNiSi)_{1-x}(Fe_2Ge)_x$, where x = 0.34 before (blue) and after (red) irradiation treatment.

## Magnetic Property Changes:

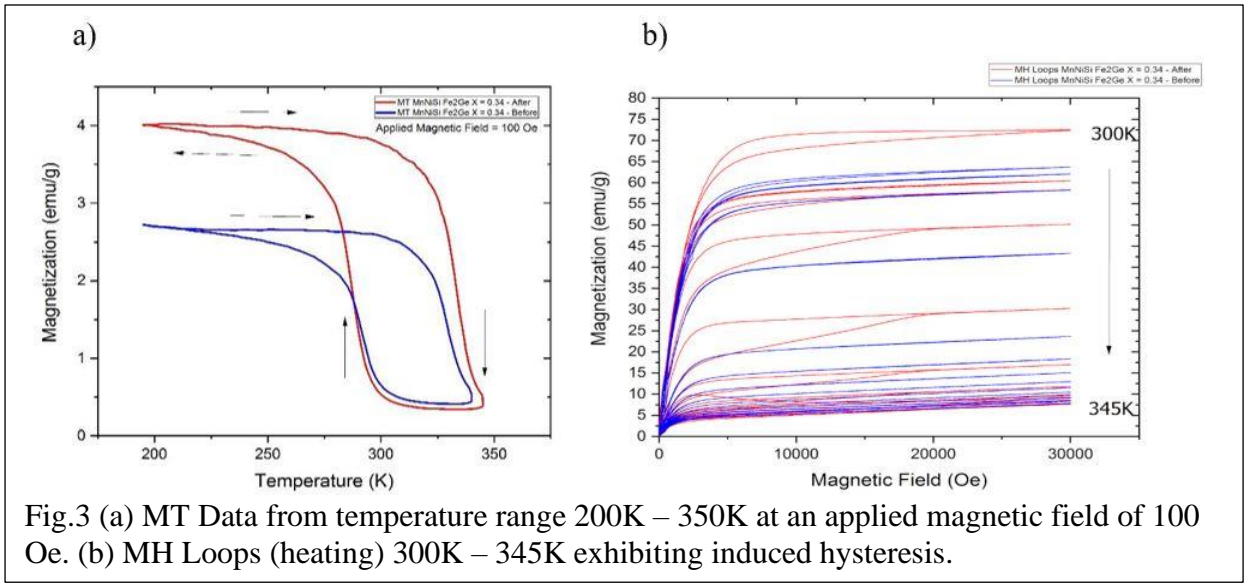

Fig.3 (a) MT Data from temperature range 200K – 350K at an applied magnetic field of 100 Oe. (b) MH Loops (heating) 300K – 345K exhibiting induced hysteresis.

Fig. 3 (a) shows the MT (magnetization vs temperature) data from 200K to 350K at an applied magnetic field of 100 Oe. The red-labeled plots are the reported magnetic properties results of the irradiated sample and the blue-labeled plots are for the pristine sample respectively. The samples exhibited first-order magneto-structural phase transition temperatures indicated by a large thermal hysteresis at the transition temperature in both the pristine and irradiated samples. The dM/dT curves (not presented in the paper) showed a first-order phase transition temperature at $T_c = \sim 292K$ for the pristine sample before irradiation treatment and at $T_c = \sim 286K$ for the irradiated sample. The phase transition is almost complete at 350K but would have continued completed at a slightly higher temperature if the measurement were conducted beyond 350K. There was also a notable change in the saturation magnetization at 100 Oe from 2.72 emu/g to 4.01 emu/g with an estimated increase of ~ 47.4% between the pristine and irradiated samples. Fig. 3 (b) presents the MH

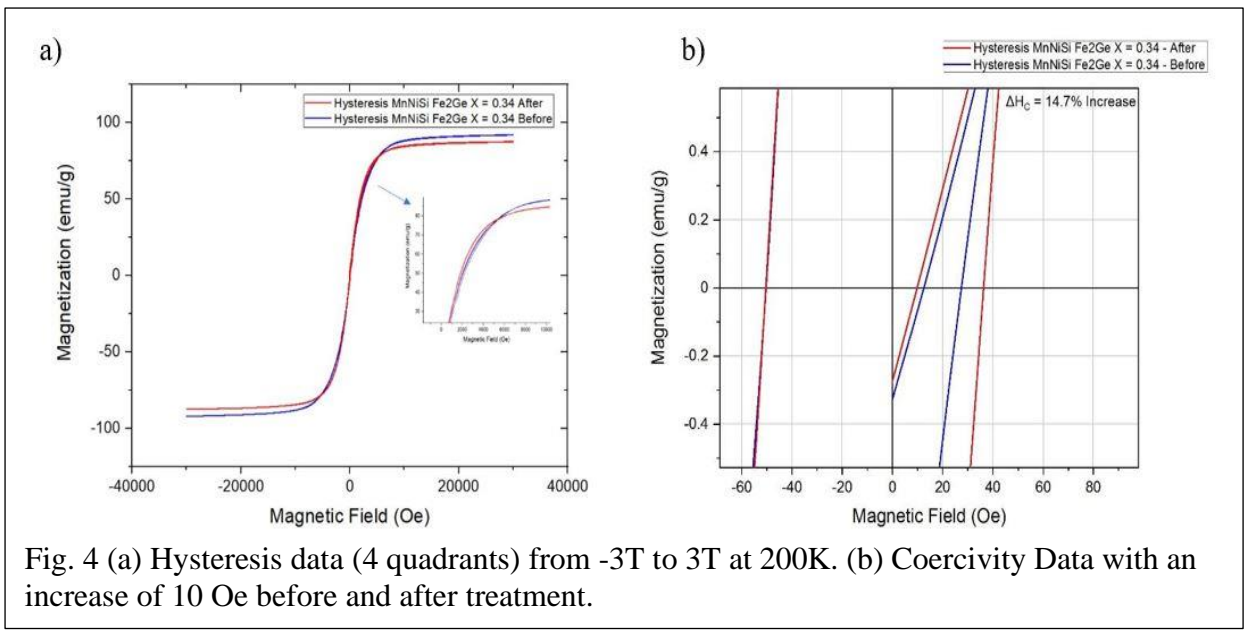

Fig. 4 (a) Hysteresis data (4 quadrants) from -3T to 3T at 200K. (b) Coercivity Data with an increase of 10 Oe before and after treatment.

(magnetization vs. magnetic field) loops from 300K to 345K with an applied magnetic field range of 0 to 30,000 Oe ($\mu_0H$= 3Tesla). It is evident from the measurement that $(MnNiSi)_{1-x}(Fe_2Ge)_x$, x = 0.34 exhibited irradiation-induced magnetic hysteresis in comparison to the pristine sample, more specifically in the range of 300K to 320K, which is near $T_c$ as observed from Fig 4(a). It can also be seen from the isotherms that the magnetization of the irradiated sample at 300K is higher than the magnetization of the pristine sample at the same temperature. This is consistent with MT measurements, where pristine samples show lower magnetization as compared to the irradiated sample. Irradiation showed to change the magnetic properties marginally. However, the selected dose didn't induce a structural or magnetic phase change in the material for this composition in both M vs. T or M vs. H measurements near the transition temperature.

Fig. 4 (a) presents the hysteresis graphs at T = 200K of the before and after 5-hour treatment. As observed in the presented results, the $(MnNiSi)_{1-x}(Fe_2Ge)_x$ sample that underwent irradiation treatment shows a small decrease in saturation magnetization. However, it can then be seen on the enlarged image that the results correspond to previously reported results that magnetization is higher for the irradiated sample before saturation than the pristine sample. T = 200 K was the temperature selected as the range is away from $T_c$ and the saturation reported in the figure is the true saturation of $(MnNiSi)_{1-x}(Fe_2Ge)_x$. The increase in the coercivity could be attributed to reversible or irreversible defect formation due to irradiation. We have noticed the sample left for several days showed recovery in magnetic property changes (not presented in the paper). This could be due to the annihilation of reversible defects (dislocation) from the room-temperature thermal energy. Further systematic studies are needed to investigate the time-dependent recovery in magnetic property changes. Fig. 4 (b) exhibits an observable change of $\Delta H_c = 14.7\%$ which revealed a 10 Oe increase in the magnetic coercivity of $(MnNiSi)_{1-x}(Fe_2Ge)_x$, x = 0.34 at 200K after irradiation treatment.

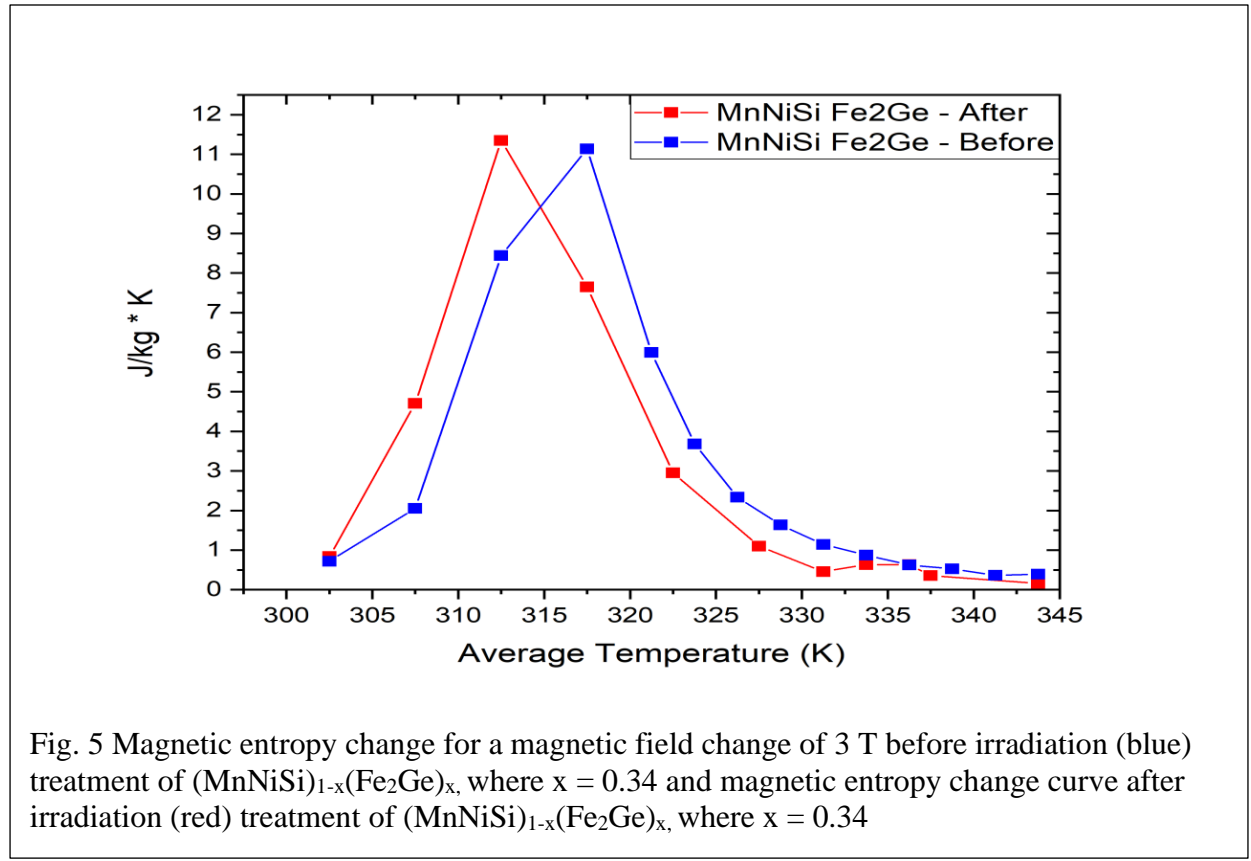

Fig. 5 Magnetic entropy change for a magnetic field change of 3 T before irradiation (blue) treatment of $(MnNiSi)_{1-x}(Fe_2Ge)_x$, where x = 0.34 and magnetic entropy change curve after irradiation (red) treatment of $(MnNiSi)_{1-x}(Fe_2Ge)_x$, where x = 0.34

Temperature-dependent magnetic entropy curves with instantaneous and applied magnetic field of $\mu_0H = 0T$ and 3T were calculated from the MH isotherms in the range of 300-350 K using Maxwell's relations for before and after irradiation-treated samples. Fig. 5 presents a maximum entropy change for the irradiated sample at $\Delta S_{mag}$ at ~ 11.349 J/kgK and a $T_{ave}$ peak of 312.5 K. Moreover, the pristine sample in Fig. 5 shows maximum entropy change at $\Delta S_{mag}$ at ~ 11.139 J/kgK and a $T_{ave}$ peak of 317.5 K. There was a slight difference in its entropy change value and a difference of 5K in the average temperature where it had its major $\Delta S_{mag}$ peak. The peak $\Delta S_{mag}$ values at the transition temperature are lower than the bulk samples reported in the literature [6] which is because of the powdered nature of our samples used for irradiation.

**Conclusions**

The powdered samples of the $(MnNiSi)_{1-x}(Fe_2Ge)_x$ with composition x=0.34 were irradiated with X-ray radiation absorbed dose up to ~35kGy. The samples' magnetic properties were characterized and showed that the irradiated samples show higher magnetization at saturation when measured between 200 to 350K with $T_c$= 292K for the pristine sample before irradiation treatment and 286K for the irradiated sample. Irradiate samples exhibited irradiation-induced magnetic hysteresis in comparison to the pristine sample in the range of 300K to 320K. The samples' magnetocaloric effect was not affected by irradiation. Isothermal entropy change, $\Delta S_{mag}$ in the temperature range of 200-350 K with an applied magnetic field change, $\Delta H$ of 3T was 11.349 J/kgK for the irradiated sample and 11.139 J/kgK for the pristine sample. Thus, this study demonstrated that prolonged exposure to multicaloric materials that undergo structural phase transition can change magnetic properties, including the phase transition temperature. Although our study did not observe changes in isothermal entropy due to irradiation, however, changes in transition temperature can affect the operating parameters such as the temperature range of the multicaloric materials. The increase in coercivity due to irradiation can induce additional heat generated from the magnetic hysteresis during the high-frequency operation. Further studies are needed to investigate if higher doses of irradiation induce magnetic or structural phase transition.

**Acknowledgments**

The authors would like to acknowledge the Dean's Undergraduate Research Initiative Fellowship – VCU College of Engineering awarded to John Peter J. Nunez and Ravi L. Hadimani.